\documentclass[acmsmall,screen]{acmart}

\AtBeginDocument{%
  }

\newcommand{\tabincell}[2]{\begin{tabular}{@{}#1@{}}#2\end{tabular}}
\usepackage{xspace}
\usepackage{url}

\usepackage[framemethod=tikz]{mdframed}
\newcounter{finding}
\newcommand{\finding}[1]{\refstepcounter{finding}
  \vspace{2.3mm}
 \begin{mdframed}[linecolor=gray,roundcorner=12pt,backgroundcolor=gray!15,linewidth=3pt,innerleftmargin=2pt, leftmargin=0cm,rightmargin=0cm,topline=false,bottomline=false,rightline = false]
  \textbf{Ans. to RQ\arabic{finding}:} #1
 \end{mdframed}
 \vspace{2.3mm}
}
\usepackage{multirow}
\usepackage{array}
\usepackage{tabularx} 
\usepackage{listings}
\usepackage{cancel}
\usepackage{enumitem}
\usepackage{threeparttable}
\usepackage{graphicx}
\usepackage{subfigure}

\mdfdefinestyle{listingstyle}{
  linewidth=0.5pt,
  linecolor=black,
  topline=true,
  bottomline=true,
  rightline=false,
  leftline=false,
  innertopmargin=5pt,
  innerbottommargin=5pt,
}

\usepackage{graphicx}
\usepackage{subcaption}
\usepackage{algorithmic}
\usepackage{graphicx}
\usepackage[ruled,vlined,linesnumbered]{algorithm2e}

\usepackage{pifont}

\usepackage{listings}
\usepackage{xcolor}

\setcopyright{acmcopyright}
\copyrightyear{2025}
\acmYear{2025}
\acmDOI{XXXXXXX.XXXXXXX}

\begin{document}




 \newcommand{\toolName}{\textit{SlsReuse}\xspace}



\title{\toolName: LLM-Powered Serverless Function Reuse}


\author{Jinfeng Wen}
\affiliation{
  \institution{Beijing University of Posts and Telecommunications}
  \city{Beijing}
  \country{China}}
\email{jinfeng.wen@bupt.edu.cn}

\author{Yuehan Sun}
\affiliation{
  \institution{Beijing University of Posts and Telecommunications}
  \city{Beijing}
  \country{China}}
\email{thesunyh2019@gmail.com}

\renewcommand{\shortauthors}{Wen et al.}

\begin{abstract}


Serverless computing has rapidly emerged as a popular cloud computing paradigm. It enables developers to implement function-level tasks, i.e., \textit{serverless functions}, without managing infrastructure. While reducing operational overhead, it poses challenges, especially for novice developers. Developing functions from scratch requires adapting to heterogeneous, platform-specific programming styles, making the process time-consuming and error-prone. Function reuse offers a promising solution to address these challenges. However, research on serverless computing lacks a dedicated approach for function recommendation. Existing techniques from traditional contexts remain insufficient due to the semantic gap between task descriptions and heterogeneous function implementations. Advances in large language models (LLMs), pre-trained on large-scale corpora, create opportunities to bridge this gap by aligning developer requirements with function semantics.


This paper presents \toolName, the first LLM-powered framework for serverless function reuse. Specifically, \toolName first constructs a reusable function repository serving as a foundational knowledge base. Then, it learns unified semantic-enhanced representations of heterogeneous functions through effective prompt engineering with few-shot prompting, capturing implicit code intent, target platforms, programming languages, and cloud services. Finally, given a natural language task query, \toolName performs intent-aware discovery combined with a multi-level pruning strategy and similarity matching. We evaluate \toolName on a curated dataset of 110 task queries. Built on ChatGPT-4o, one of the most representative LLMs, \toolName achieves Recall@10 of 91.20\%, exceeding the state-of-the-art baseline by 24.53 percentage points. Experiments with Llama-3.1 (405B) Instruct Turbo, Gemini 2.0 Flash, and DeepSeek V3 further confirm the generalization capability of \toolName, yielding consistently strong effectiveness across recent LLMs.

\end{abstract}



\maketitle

\section{Introduction}\label{sec:introduction}

Serverless computing has rapidly emerged as a prominent paradigm in cloud computing, allowing developers to concentrate on application business logic while offloading infrastructure provisioning, scaling, and maintenance to cloud providers~\cite{wen2023rise}. This paradigm has been applied across diverse domains, including machine learning~\cite{wang2019distributed, yu2021gillis}, numerical computing~\cite{shankar2020serverless}, video processing~\cite{fouladi2017encoding, ao2018sprocket}, Internet of Things~\cite{zhang2021edge, jindal2021function}, and big data analytics~\cite{gimenez2019framework, enes2020real}. Reflecting its increasing adoption and economic impact, the global serverless computing market is projected to grow from about $\$$19 billion in 2024 to nearly $\$$41 billion by 2028~\cite{marketreport}.


In serverless computing, developers primarily implement event-driven units of execution at the function granularity, referred to as \textit{serverless functions}, which interact with diverse cloud services to accomplish specific tasks. These functions run on fully managed serverless platforms, such as AWS Lambda~\cite{aws}, Google Cloud Functions~\cite{google}, and Azure Functions~\cite{azure}, achieving high elasticity and operational simplicity without direct server management. Despite these advantages, serverless computing presents considerable barriers, particularly for novice developers~\cite{Wen21challenges}. In practice, developers are required to implement serverless functions from scratch to fulfill specific functionality requirements. This process is time-consuming, cognitively demanding, and error-prone, owing to the novel programming paradigm and the style heterogeneity of serverless platforms~\cite{Wen21challenges}. Each platform has its own language support, function interfaces, and service configuration schemes, forcing developers to consult extensive official documentation to understand details~\cite{wen2021characterizing}. The absence of unified standards and comprehensive guidance~\cite{Wen21challenges} further steepens the learning curve. As discussed~\cite{hardserverless}, high-quality tutorials and guides remain scarce, making it difficult for developers to access reliable information when learning serverless technologies. Consequently, developing serverless functions from scratch imposes a substantial burden on developers.


One promising solution is to reuse proven serverless functions developed by other developers. However, in the serverless computing research, the lack of discovery and recommendation mechanisms hinders developers from leveraging reusable functions. Realizing function recommendation requires addressing several challenges. (1) Lack of reusable function repositories: Unlike microservices, serverless functions currently lack well-structured, dedicated repositories that facilitate function sharing and reuse. (2) Function code heterogeneity: Serverless function implementations exhibit significant variation not only in task objectives but also in underlying platforms, programming languages, and cloud service integrations. This heterogeneity hinders consistent interpretation and comparison. (3) Function and requirement gap: With the new paradigm of serverless computing, developers generally express task requirements in natural language, which tends to be unstructured, ambiguous, and imprecise. These descriptions diverge from the terminology and abstractions used in function implementations, creating a gap that hinders accurate function recommendation.


Existing code recommendation approaches can be broadly categorized into code-to-code and text-to-code approaches. 
Code-to-code recommendation methods~\cite{kim2018facoy, luan2019aroma, silavong2022senatus, nguyen2020code} discover relevant code snippets based on partial code inputs. While effective in code-centric scenarios, they are fundamentally misaligned with function reuse driven by text-based task requirements.
In contrast, text-to-code recommendation approaches~\cite{chatterjee2009sniff, cambronero2019deep, gu2018deep, sachdev2018retrieval, gao2023know} map natural language descriptions to code snippets, by treating code as a bag of words and applying keyword-based search~\cite {chatterjee2009sniff} or embedding representation calculation~\cite {cambronero2019deep, gu2018deep, sachdev2018retrieval}. However, these methods overlook high-level development goals and fail to capture implicit functional intent, as code is reduced to surface-level tokens. Moreover, essential programming characteristics in serverless contexts, such as the chosen serverless platform and cloud service integrations, are not incorporated. These factors are critical for accurate recommendations and effective function development in serverless computing. Finally, code tokens used in existing approaches remain misaligned with natural language requirements, as the two operate at different levels of abstraction. Overall, existing methods lack the useful semantic modeling and intent alignment required to support effective serverless function reuse.



Recent advancements in Large Language Models (LLMs) offer a prospective direction for addressing the challenges of serverless function reuse. LLMs have achieved impressive performance in a variety of software engineering tasks, such as program repair~\cite{fan2023automated}, unit test generation~\cite{yuan2024evaluating}, and log parsing~\cite{xu2024divlog}. Trained on massive amounts of publicly available data, including source code and natural language text, LLMs are capable of understanding code implementations, extracting key information, and inferring task intent. These capabilities make LLMs particularly well-suited to enable effective serverless function reuse.


In this paper, we introduce \toolName, the first framework for serverless function reuse, which leverages LLMs to enhance semantic representations. \toolName tackles key challenges in serverless computing, including the lack of reusable functions, heterogeneity in function implementations, and the semantic gap between task descriptions and serverless functions, through a combination of structured knowledge extraction, semantic representation, and intent-aware recommendation. 
Specifically, \toolName first builds a reusable serverless function repository by acquiring functions from open-sourced benchmarks and applying quality-driven filtering. This repository forms the foundational base for function recommendations. Second, \toolName introduces a semantic representation paradigm that abstracts function code into a unified structured space, capturing code intent, target platform, programming languages, and cloud services. This leverages few-shot prompt engineering of LLMs to extract consistent semantic knowledge across heterogeneous implementations. Finally, given a task query, \toolName performs intent-aware function discovery and recommendation. The query is semantically encoded via LLMs and matched against precomputed function representations. A multi-level pruning strategy removes functions with inconsistent attributes, while similarity-driven matching ranks candidate functions according to alignment with task intent.

To evaluate \toolName, we construct a reusable repository of 500 serverless functions spanning diverse domains and programming languages, as no public dataset is available. We further create 110 task queries to support evaluation. Results show that \toolName, powered by ChatGPT-4o  (one of the most representative LLMs known for outstanding performance), achieves \textit{Recall@1}, \textit{@5}, \textit{@10}, \textit{@15}, and \textit{@20} of 52.13\%, 86.13\%, 91.20\%, 92.67\%, and 92.93\%, exceeding the state-of-the-art baseline by 28.80, 35.47, 24.53, 18.67, and 15.60 percentage points, respectively. In terms of ranking quality measured by Mean Reciprocal Rank (MRR), \toolName reaches \textit{MRR@1} of 0.5240, \textit{MRR@5} of 0.6547, \textit{MRR@10} of 0.6616, \textit{MRR@15} of 0.6628, and \textit{MRR@20} of 0.6629, yielding relative improvements of 124.57\%, 93.19\%, 84.15\%, 81.58\%, and 80.70\% over the state-of-the-art baseline, respectively. Compared with a customized LLM-based variant method, our multi-level pruning strategy reduces average recommendation latency by 44.36\%. 
In addition, we test with alternative LLM backends, including Llama 3.1 (405B) Instruct Turbo, Gemini 2.0 Flash, and DeepSeek V3, to evaluate generalization. Results confirm the generalization capability of \toolName and consistently high effectiveness across all models.


In summary, this paper makes the following contributions:

\begin{itemize}[leftmargin=*]
\item We present \toolName, the first framework for serverless function reuse. It introduces a semantic-enhanced representation that uses LLMs with designed few-shot prompts to extract knowledge.




\item We conduct an empirical study on this dataset to evaluate the effectiveness of \toolName, showing that it consistently outperforms state-of-the-art baselines.
\end{itemize}

\section{Background on Serverless Computing}\label{sec:background}

\subsection{Difference between Serverless Computing and Other Paradigms}


Serverless computing, commonly implemented through Function-as-a-Service (FaaS) serverless platforms such as AWS Lambda~\cite{aws}, Google Cloud Functions~\cite{google}, and Azure Functions~\cite{azure}, allows developers to deploy lightweight, stateless functions. These functions are automatically triggered by external events, including HTTP requests, message queues, and file uploads. Unlike traditional monolithic applications, serverless functions are designed to be modular, single-purpose, and ephemeral, executing within fully managed cloud environments. While both serverless functions and microservices emphasize modularity and decoupling, they differ fundamentally. Microservices are long-running, stateful services that expose APIs and maintain dedicated resources. In contrast, serverless functions are stateless, event-driven, and invoked on demand, with infrastructure management fully handled by the serverless platform.

\subsection{Development Process of Serverless Functions}


\begin{figure*}[t]
	\centering
    \includegraphics[width=0.98\textwidth]{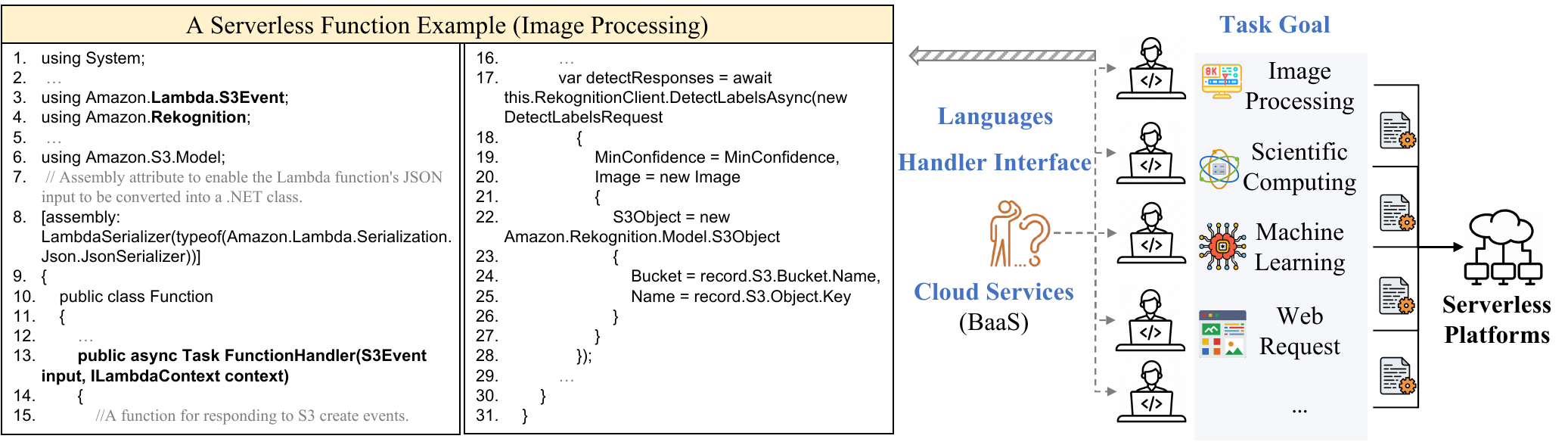}
    \caption{The development process and an example of a serverless function.}
    \label{fig:functiondevelopment}
\end{figure*}

As illustrated in Fig.~\ref{fig:functiondevelopment}, implementing a serverless function involves encapsulating the intended task goal (e.g., image processing) in a platform-supported programming language and adhering to the handler interface defined by the target platform. A key feature of this process is the seamless integration with cloud services, referred to as Backend-as-a-Service (BaaS). The function can interact with fully managed services such as object storage (e.g., AWS S3), analysis service (e.g., AWS Rekognition), and messaging systems (e.g., Google Pub/Sub). These services are accessed via SDKs or RESTful APIs, allowing functions to perform operations such as file processing and database access without re-implementing complex backend functionality. The function finally is deployed and executed with configuration metadata that specifies its runtime parameters (e.g., memory allocation size) and associated event triggers.

Fig.~\ref{fig:functiondevelopment} illustrates a serverless function implemented in C\# and deployed on AWS Lambda, sourced from the official AWS GitHub repository \cite{serverlessfunctioncode}. The function is triggered by AWS S3 object creation events (line 3). Upon invocation through the C\# handler interface of AWS Lambda (line 12), it first verifies whether the uploaded object is an image. If the condition holds, the function calls cloud service AWS Rekognition to perform image label detection and subsequently appends the detected labels as metadata tags to the corresponding S3 object (lines 17–29).

While serverless computing abstracts infrastructure management and facilitates rapid, event-driven function development, it simultaneously introduces complexity in function design and cloud service integration. Thus, reusing serverless functions is essential to improve development efficiency, reduce code redundancy, and sustain a maintainable cloud function ecosystem.

\section{\toolName: our serverless function reuse framework}\label{sec:tool}

\begin{figure*}[t]
	\centering
    \includegraphics[width=0.98\textwidth]{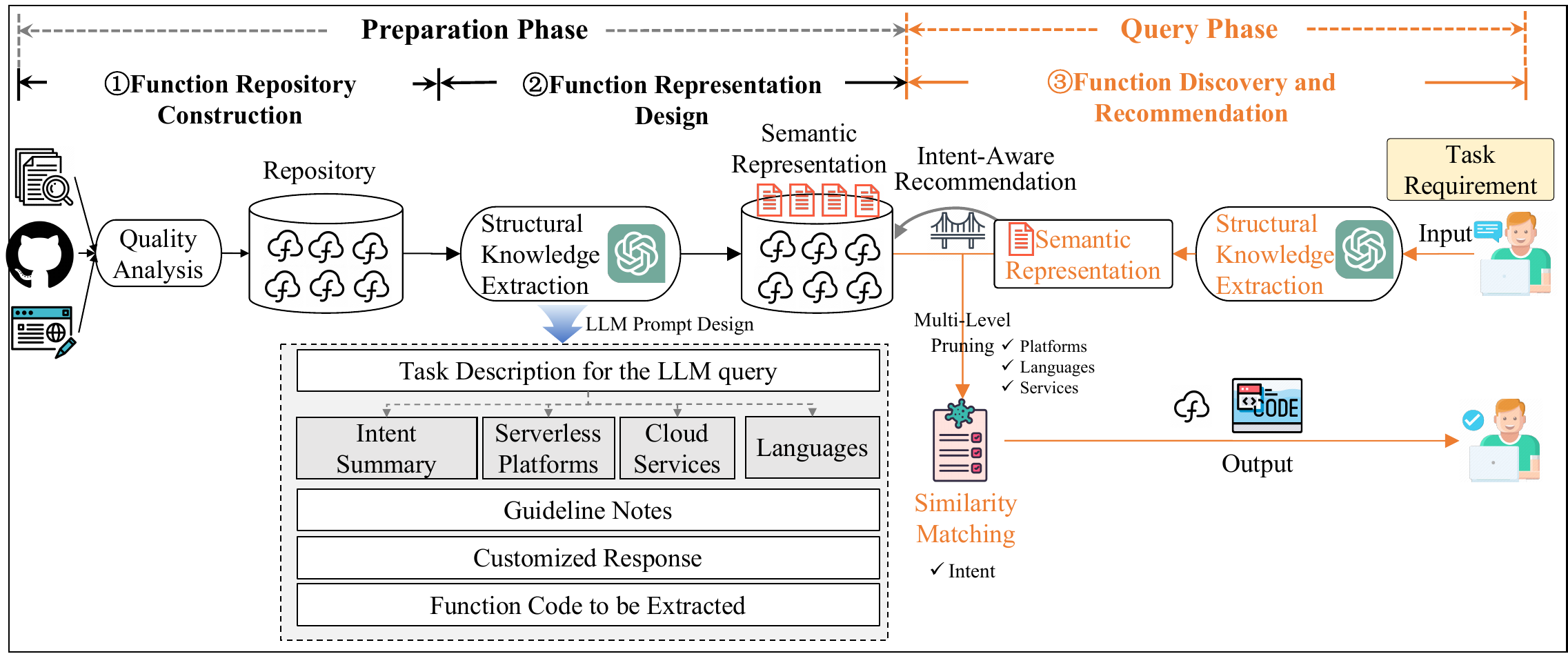}
    \caption{The overview of \toolName.}
    \label{fig:overview}
\end{figure*}



We present \toolName, a semantic-enhanced framework for serverless function reuse. It addresses the challenges of recommending heterogeneous functions through three contributions. First, it alleviates the scarcity of reusable functions by constructing a curated, high-quality repository. Second, it enables consistent interpretation and comparison by abstracting diverse code into a unified, semantics-enriched representation space. Third, it narrows the gap between natural language queries and serverless functions via intent-aware discovery.

Fig.~\ref{fig:overview} illustrates the overall architecture of \toolName, which is organized into two phases: the preparation phase and the query phase, and encompasses three core components. \textcircled{1} \textbf{Function Repository Construction} curates a high-quality repository of serverless functions from open-source ecosystems (e.g., GitHub). Through quality analysis, it ensures the collected functions are reliable and reusable, establishing the data foundation for function reuse. \textcircled{2} \textbf{Function Representation Design} introduces a novel semantic representation for serverless functions. It employs a tailored LLM prompt to conduct structural knowledge extraction, abstracting code into a unified semantic space. The resulting enriched representations enable consistent interpretation and comparison across heterogeneous implementations. \textcircled{3} \textbf{Function Discovery and Recommendation} aligns developer task requirements with repository functions by transforming natural language inputs into semantic representations using structural knowledge extraction. It then applies multi-level pruning and similarity matching to identify and recommend functions with aligned intent. 
Next, we describe these components in detail.



\subsection{Function Repository Construction}\label{sec:functionconstruction}

This component is dedicated to constructing a diverse, scalable, and high-quality repository of serverless functions, which serves as the foundation for function recommendation. To this end, we curate functions from publicly available and widely adopted open-source benchmarks. These functions are well-tested and suitable for reuse. Our repository aims to span a wide range of application domains (e.g., image processing, machine learning, Web services), serverless platforms (e.g., AWS Lambda, Azure Functions), and programming languages (e.g., Python, JavaScript, C\#).

To ensure the functional quality of the collected functions, we design a filtering and normalization process: (1) \textit{Trivial function elimination.} We apply heuristic-based filtering to exclude non-substantive or boilerplate functions. Specifically, we remove ``Hello World''-style examples, placeholder code identified through manual inspection, and structurally trivial functions characterized by minimal logic or return-only statements. 
(2) \textit{Benchmark script filtering.} To isolate reusable functions from evaluation-related scripts, we further exclude benchmark-specific measurement or orchestration code that is not part of the serverless function's core task logic. (3) \textit{Function management normalization.} Each retained serverless function is encapsulated as a uniquely identified unit containing all relevant code. These units constitute the repository's fundamental entities.




To accommodate the dynamic and evolving nature of the serverless ecosystem, this component provides a batch processing interface that supports the incremental integration of new functions into the repository. This design ensures that the repository remains up to date, comprehensive, and well-suited for downstream analysis and recommendation.


\subsection{Function Representation Design}


To interpret and compare heterogeneous serverless functions, this component introduces a new semantic representation paradigm. Unlike prior work based on keywords or embeddings~\cite{chatterjee2009sniff, cambronero2019deep, gu2018deep, sachdev2018retrieval}, our design captures higher-level semantics tailored to serverless functions, encompassing both explicit development characteristics and implicit task-level intent.

Given the diversity of programming styles, the scarcity of metadata, and the implicit semantics embedded in code logic, direct information extraction remains challenging. To overcome them, we leverage the code reasoning and comprehension capabilities of LLMs to extract key information and thereby enhance the semantics of each function. Recent studies~\cite{lian2023configuration, yuan2024evaluating} have demonstrated that \textit{Prompt Engineering} can effectively enhance LLM performance across various tasks. It involves designing task-specific instructions (i.e., prompts) to guide model behavior without modifying parameters. Through carefully crafted prompts, LLMs can thus adapt to a wide range of tasks.

\begin{figure*}[t]
	\centering
    \includegraphics[width=0.98\textwidth]{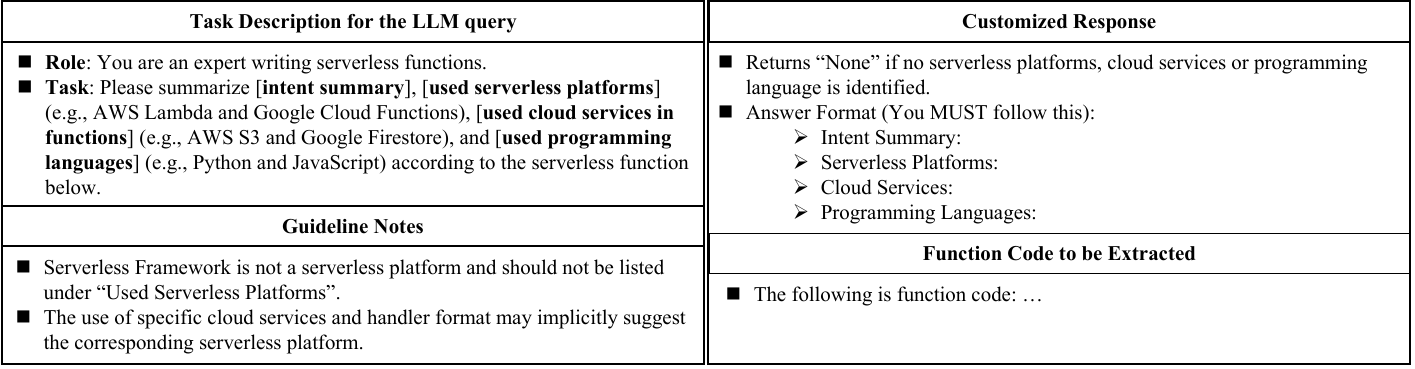}
    \caption{The prompt of structured knowledge extraction used in \toolName.}
    \label{fig:promptexample}
\end{figure*}

Building on this, we design a structured prompt to guide the LLM in extracting critical semantic attributes from raw function code. By coupling the prompt with a small set of curated examples (i.e., few-shot learning~\cite{lian2023configuration, yin2024multitask, xu2024divlog}), the LLM can infer and summarize latent information according to examples. Our prompt consists of four parts: (1) task description for the LLM query, (2) guideline notes, (3) customized response, and (4) function code to be extracted. The overall structure is illustrated in Fig.\ref{fig:overview}, with a concrete example shown in Fig.\ref{fig:promptexample}. We elaborate on each part below.

$\bullet$ The task description for the LLM query part includes both a role-play setting and a detailed task instruction. The role-play instruction (e.g., ``You are an expert writing serverless functions'') is a widely used prompt optimization technique~\cite{yuan2024evaluating} to improve output quality. The task instruction explicitly specifies the extraction of attribute information. As illustrated earlier in Fig.~\ref{fig:functiondevelopment}, serverless functions may implement different task goals across various languages, handler interfaces tailored to the target platform, and supported cloud services. These factors are critical to function development and thus constitute the key extraction targets: (1) an intent summary of the serverless function, (2) the serverless platforms used, (3) the cloud services employed, and (4) the programming languages adopted. Moreover, the prompt embeds domain-specific examples (i.e., few-shot learning) within the task instruction. For instance, serverless platform examples include AWS Lambda and Google Cloud Functions; cloud service examples include AWS S3 and Google Firestore; and programming language examples include Python and JavaScript.


$\bullet$ The guideline notes part is designed to help the LLM resolve ambiguities and capture implicit associations during information extraction. They are distilled from common misclassification patterns observed in our function analysis. First, the note clarifies that the ``Serverless Framework'' is a development framework rather than a serverless platform, a distinction often confused due to the name similarity. Second, it emphasizes that the use of certain cloud services and specific handler interfaces generally implies the adoption of a corresponding serverless platform, even if it is not explicitly declared in the code. These notes ensure more accurate extraction.


$\bullet$ The customized response part defines both the structural and content constraints for the LLM's output. To avoid vague or overly general responses, the LLM is instructed to return ``None'' when no relevant serverless platform, cloud service, or programming language can be identified. When such information is available, it must be organized in a standardized four-part format: Intent Summary, Serverless Platforms, Cloud Services, and Programming Languages. This output ensures representational consistency and simplifies subsequent information analysis.


After applying the prompt-based knowledge extraction, the structural semantic information of each serverless function is obtained. However, identical concepts may appear under different lexical forms across functions. For instance, the extracted programming language ``JavaScript'' may be represented as ``JS''. To address such inconsistencies, a mapping table is constructed to normalize terminology. Finally, we organize the extracted information as a quadruple semantic representation \textbf{\{intent content, serverless platforms, cloud services, programming languages\}}. 


A concrete semantic representation of the function in Fig.~\ref{fig:functiondevelopment} is illustrated. Its intent can be summarized via ChatGPT-4o as follows: \textit{The code defines a serverless function that responds to S3 create events. When an image file is uploaded to an S3 bucket, the function uses Amazon Rekognition to detect labels in the image. It then tags the S3 object with the detected labels and their confidence scores. The function includes an integration test to ensure that the image is processed correctly and tagged with labels.} To enable subsequent content similarity analysis, the intent summary is encoded into a high-dimensional vector representation using sentence embedding models such as Sentence-BERT (default dimensionality: 384). For example, this summary is transformed into a vector [-0.0411, -0.0280, ..., 0.0068, -0.0334]. In addition, the extracted attributes of the function in Fig.~\ref{fig:functiondevelopment} include the serverless platform (\textit{AWS Lambda}), the associated cloud services (\textit{AWS S3} and \textit{AWS Rekognition}), and the programming language (\textit{C\#}). These results also further suggest that LLMs can perform a relatively effective information extraction. The final semantic-enhanced representation is therefore expressed as \{[-0.0411, -0.0280, ..., 0.0068, -0.0334], AWS Lambda, [AWS S3, AWS Rekognition], C\#\}.

\subsection{Function Discovery and Recommendation}


After the preparation phases, we enter the query phase, where relevant serverless functions are recommended based on a given task query. We employ an intent-aware discovery and recommendation strategy. The procedure is outlined in Algorithm~\ref{alg:functiondiscovery}, which comprises four main steps.

\begin{algorithm}[t]
\caption{Intent-Aware Function Discovery and Recommendation}
\label{alg:functiondiscovery}
\KwIn{Task query $Q$ in natural language}
\KwOut{Top-$k$ recommended serverless functions $\{F_1, F_2, \dots, F_k\}$}

\textbf{Step 1: Query Understanding} \\
Extract attribute values from $Q$: task intent summary $T_Q$, serverless platforms $P_Q$, cloud services $C_Q$, programming languages $L_Q$ \\
Form the quadruple semantic representation of the query $V_Q = \{T_Q, P_Q, C_Q, L_Q\}$

Initialize full function set $\mathcal{F}_0 \leftarrow$ all semantic representations of functions in repository

\textbf{Step 2: Multi-level Pruning} \\
\ForEach{attribute type $A \in \{P, C, L\}$}{
    Initialize filtered candidate set $\mathcal{F}_A \leftarrow \emptyset$ \\
    Initialize exact match set $\mathcal{F}^{\text{full}}_A \leftarrow \emptyset$ \\
    Initialize partial match set $\mathcal{F}^{\text{partial}}_A \leftarrow \emptyset$ \\

    \If{$A_Q$ $\neq$ \text{None}}{  \tcp{Determine whether current attribute of the query is not ``None''.}
    
    \ForEach{$F_i \in \mathcal{F}_0$}{
        Extract attribute set $A_i$ of $F_i$ \\
        Compute Jaccard Distance: $J_A = 1 - \frac{|A_Q \cap A_i|}{|A_Q \cup A_i|}$ \\
        Compute Subset Coverage: $S_A = \frac{|A_Q \cap A_i|}{|A_Q|}$ \\
        
        \eIf{$S_A = 1$}{
            Add $F_i$ to $\mathcal{F}^{\text{full}}_A$  \tcp{Complete or Superset Match}
        }{
            Record objective vector $M^A_i = \{J_A, 1 - S_A\}$ \\
            Add $(F_i, M^A_i)$ to $\mathcal{F}^{\text{partial}}_A$
        }
    }

    Apply Pareto Optimization over all $M^A_i$ in $\mathcal{F}^{\text{partial}}_A$ \\
    Let $\mathcal{F}^{\text{pareto}}_A \leftarrow$ Pareto-optimal functions \\
    
    Merge: $\mathcal{F}_A \leftarrow \mathcal{F}^{\text{full}}_A \cup \mathcal{F}^{\text{pareto}}_A$ \\
    Update: $\mathcal{F}_0 \leftarrow \mathcal{F}_A$
    }
}

\textbf{Step 3: Intent Similarity Calculation} \\
\ForEach{$F_i \in \mathcal{F}_0$}{
    Encode: $E_Q = \text{Embed}(T_Q)$, $E_i = \text{Embed}(T_i)$ using Sentence-BERT \\
    Compute cosine similarity: $S_i = \cos(E_Q, E_i)$
}

\textbf{Step 4: Top-k Recommendation} \\
Sort $\mathcal{F}_0$ by descending $S_i$ \\
Return top-$k$ serverless functions from the ranked list

\end{algorithm}

\textit{\textbf{Step 1: Query Understanding.}} Given a task query, we first derive its semantic representation using a prompt-based structural knowledge extraction analogous to that employed in the preparation phase for serverless functions (line 2). Specifically, we obtain the key information of the query: task intent summary $T_Q$, target platforms $P_Q$, desired cloud services $C_Q$, and programming languages $L_Q$. $V_Q = \{T_Q, P_Q, C_Q, L_Q\}$ will establish a link between the high-level task query $Q$ and the low-level function implementations $\mathcal{F}_0$ (lines 2-3). 

Considering a task query: \textit{The generated function handler responds to S3 events on an Amazon S3 bucket and if the object is a png or jpg file uses Amazon Rekognition to detect labels. Once the labels are found it adds them as tags to the S3 Object.} 
This query description is extracted from the GitHub README~\cite{serverlessfunctioncode} of the function in Fig.~\ref{fig:functiondevelopment}. For this example, ChatGPT-4o infers the serverless platform $P_Q$ as \textit{AWS Lambda}, the cloud services $C_Q$ as \textit{AWS S3} and \textit{AWS Rekognition}, while the programming language is labeled as ``None'' due to the absence of explicit language-specific information.



\textit{\textbf{Step 2: Multi-level Pruning.}} To improve the efficiency of function discovery, irrelevant functions are first pruned based on task requirements, yielding a set of relevant candidate functions for subsequent recommendation. Our pruning process leverages extracted structural attributes for filtering. Specifically, we analyze the query representation $V_Q$ and function representations $\mathcal{F}_0$. We first narrow the search space to functions sharing the same fundamental development attributes, e.g., serverless platforms $P$, cloud services $C$, and languages $L$, as specified in the query (line 6). For each attribute type $A$ $\in \{P, C, L\}$, if the corresponding value in the task query $Q$ is not ``None'', candidate functions are matched accordingly (lines 10-23); otherwise, the procedure proceeds to the next attribute type. This forms a \textbf{multi-level pruning strategy over attribute types}. Ultimately, only functions with compatible development characteristics are retained for subsequent ranking.

We next introduce the process of generating candidate functions based on the information associated with the current attribute type.
At this process, a key challenge arises: attribute values in a task query may contain multiple elements, such as several cloud services (e.g., \textit{Amazon S3} and \textit{Amazon Rekognition} in the example query). Functions in the repository may likewise integrate multiple attribute values. Consequently, strict one-to-one matching is overly restrictive, while fixed threshold-based filtering of collapsing all factors into a single weighted score risks discarding partially relevant candidates. To address this, we formulate the relevant function discovery on the current attribute type as a multi-objective optimization problem and, building on Pareto optimization~\cite{qian2015subset}, propose \textbf{a customized Pareto-based function selection method} tailored to our scenario. Rather than setting a fixed, pre-defined threshold for filtering, Pareto optimization retains all non-dominated candidates, i.e., functions for which no other candidate performs strictly better across all defined optimization metrics. This way preserves a diverse set of Pareto-optimal functions. To this end, we define two complementary optimization metrics: \textbf{symmetric and asymmetric} (lines 13-14). The symmetric metric, \textit{Jaccard Distance}, measures the overall similarity between the query and a candidate function based on the overlap of elements within the current attribute type. It is computed as $J_A = 1 - \frac{|A_Q \cap A_i|}{|A_Q \cup A_i|}$. The value ranges from 0 to 1, where 0 indicates complete overlap (maximum similarity) and 1 indicates no overlap (maximum dissimilarity). Owing to its symmetry, it is well-suited for measuring holistic overlap. The asymmetric metric, \textit{Subset Coverage}, quantifies the proportion of attribute elements of the task query satisfied by the candidate function. It is computed as $S_A = \frac{|A_Q \cap A_i|}{|A_Q|}$. The value also ranges from 0 to 1, where 1 indicates full coverage of the query attributes (optimal), and 0 indicates no coverage. This is particularly important for verifying whether a function meets core development requirements, even if it contains additional, potentially irrelevant elements. Note that $A_Q$ denotes the set of elements for the given attribute type in the task query, and $A_i$ represents the corresponding set for function $F_i$. The reason for using these two metrics is: \textit{Jaccard Distance} may fail to capture the inclusion of essential elements, while \textit{Subset Coverage} may tolerate extraneous elements that introduce overhead. 
Integrating both within the Pareto optimization process enables a more balanced and nuanced evaluation of candidate functions.
Based on the defined optimization metrics, potential candidate functions are selected. First, to prioritize candidates that fully satisfy the task query, we retain all functions with complete subset coverage (\textit{Subset Coverage} = 1), including both exact and superset matches, denoted as $\mathcal{F}^{\text{full}}_A$ (lines 15-16). These functions fully implement the required elements and are directly added to the candidate set. Second, for the remaining functions with partial coverage, our Pareto-based function selection is applied using the two defined metrics to identify non-dominated candidates, denoted as $\mathcal{F}^{\text{pareto}}_A$, those representing optimal trade-offs between matching completeness and specificity (lines 17-21). Finally, we merge $\mathcal{F}^{\text{full}}_A$ and $\mathcal{F}^{\text{pareto}}_A$ to obtain candidate set $\mathcal{F}_A$ (line 22).

This generation candidate function process is performed independently for each attribute type (platforms $P$, services $C$, languages $L$). At each stage, only the retained subset $\mathcal{F}_A$ (which updates $\mathcal{F}_0$) is forwarded to the next attribute filter, progressively refining the candidate functions (line 23). Through such a multi-level pruning, the resulting set ensures both development compatibility and requirement satisfaction. In our provided task query example, this process ultimately selects 33 candidate functions from a repository of 500 functions, as further detailed in Section~\ref{sec:functionrepository}.


\textit{\textbf{Step 3: Intent Similarity Calculation.}} We further refine the candidate set $\mathcal{F}_0$ by evaluating intent similarity using extracted task intent summary. 
Since the summary is expressed in natural language and cannot be directly compared via string matching, we adopt Sentence-BERT to encode intent descriptions into high-dimensional vectors (line 26). Then, we compute cosine similarity between the task query's intent vector and each function's intent vector (line 27). Cosine similarity~\cite{cosinesimilarity} is a standard metric for vector comparison. In the task query example, the similarity between its intent summary and that of the function code in Fig.~\ref{fig:functiondevelopment} is 0.9001, indicating high intent relevance.


\textit{\textbf{Step 4: Top-$k$ Recommendation.}} We recommend the top-$k$ functions with the highest intent similarity scores, providing a ranked list of candidates that best match their task requirements. In the task query example, the function shown in Fig.~\ref{fig:functiondevelopment} is ranked first. In fact, this function is also the correct function corresponding to the provided task query.

\underline{\textit{Complexity analysis of Algorithm~\ref{alg:functiondiscovery}.}} Let $n$ denote the number of functions in the repository, $m$ the number of attribute types (a small constant, e.g., 3), $d$ the average number of elements per function per attribute type, and $l$ the dimensionality of the intent vectors (e.g., 384 for Sentence-BERT). Step 1 incurs negligible cost with constant-time query parsing. In Step 2, for each of the constant $m$ attribute types, the algorithm computes \textit{Jaccard Distance} and \textit{Subset Coverage} for $n$ functions in $O(nd)$, followed by Pareto optimization, which requires $O(n^2)$ in the worst case due to pairwise dominance checks. Thus, this step dominates with $O(n^2)$ time complexity. Step 3 performs similarity calculation for each function (in the worst-case scenario where all $n$ functions are retained as candidates) and the query, resulting in a cost of $O(nl)$. Step 4 sorts the results in $O(nlogn)$. Overall, the total time complexity is $O(n^2 + nl + nlogn)$, which simplifies to $O(n^2)$ as the dominant term. For the space complexity, the algorithm stores function attributes ($O(nd)$), intent vectors ($O(nl)$), and Pareto vectors ($O(mn)$), leading to total space complexity $O(n(d + l + m))$, which reduces $O(nl)$ assuming $d$ and $m$ are constants. Overall, Algorithm~\ref{alg:functiondiscovery} is quadratic in time and linear in space with respect to the number of functions.

\subsection{Implementation}
We implement \toolName as a Python-based prototype. The design is language- and platform-agnostic, making it applicable to serverless function analysis across diverse programming languages and serverless platforms, as well as to function recommendation for arbitrary task queries. The implementation is fully encapsulated within the tool and remains transparent to developers. In practice, developers only need to provide requirement descriptions through the tool interface, without additional effort. For each task query, \toolName generated a ranked list of relevant serverless functions, each linked directly to its original implementation. The core analysis is built upon and enhanced by advanced libraries, including OpenAI API, Scikit-learn, and Sentence Transformers.


\section{Experimental evaluation}\label{sec:evaluation}


\subsection{Research Questions}

\noindent $\bullet$ \textbf{RQ1:} How does the semantic-enhanced representation of \toolName perform compared with traditional methods?





\noindent $\bullet$ \textbf{RQ2:} What is the advantage of \toolName in leveraging multi-level pruning for recommendation?



\noindent $\bullet$ \textbf{RQ3:} How does the non-determinism of LLMs influence the effectiveness of \toolName?



\noindent $\bullet$ \textbf{RQ4:} How well does \toolName generalize when applied with different LLMs?



\subsection{Reusable Serverless Function Dataset}\label{sec:functionrepository}

Given the lack of reusable functions in current serverless ecosystems, we construct a dedicated dataset following the construction principles outlined in Section~\ref{sec:functionconstruction}. We first survey widely adopted open-source benchmarks from both academia~\cite{yu2020characterizing, kim2019functionbench, maissen2020faasdom} and industry~\cite{awsserverlessrepo}, including \textit{FunctionBench}\cite{kim2019functionbench}, \textit{ServerlessBench}\cite{yu2020characterizing}, \textit{AWS Samples}\cite{AWSSamples}, \textit{SeBS}\cite{copik2021sebs}, and \textit{FaaSDom}~\cite{maissen2020faasdom}. Based on these sources, we apply our filtering process to curate a final set of 500 high-quality serverless functions spanning diverse domains, including Web request handling, video processing, scientific computing, machine learning, and natural language processing. This dataset size is sufficient as it is comparable to, or even larger than, datasets employed in prior code search~\cite{luan2019aroma, ma2024compositional}. Moreover, our dataset encompasses functions from major serverless platforms (AWS Lambda, Google Cloud Functions, Microsoft Azure Functions, and Apache OpenWhisk) and supports a broad set of programming languages, including Python, Ruby, JavaScript, Java, C++, C\#, and Go. This language diversity not only mirrors the heterogeneity of real-world serverless ecosystems but also surpasses the scope of language support reported in prior works~\cite{luan2019aroma, silavong2022senatus, chatterjee2009sniff, raghothaman2016swim, rahman2016rack}.

\subsection{Compared Baselines}

In the absence of established approaches for serverless function reuse, we evaluate \toolName against two representative traditional methods adapted for function recommendation. We also introduce a customized LLM-based variant as a baseline to explore the additional advantages of \toolName.


$\bullet$ \textit{Keyword-based method}: SNIFF~\cite{chatterjee2009sniff} is a classical code search approach that supports free-form natural language queries about programming tasks. It processes the query and code by applying stop-word removal and stemming, treating them as a bag-of-words. Following this core idea, we adapt it to the serverless function recommendation: both the task query and the function code are preprocessed identically (stop-word removal and stemming), and candidate functions are retrieved if they contain all query keywords. As exact matches across all query keywords are uncommon, we relax this constraint and rank candidate functions in descending order of keyword match counts.



$\bullet$ \textit{Embedding-based method}: 
UNIF~\cite{cambronero2019deep} is a leading text-to-code embedding method, which directly maps queries and code into dense embedding vectors. Code fragments semantically related to a query are discovered by computing similarity between their vectors. In our adaptation, the task query is encoded into a dense embedding vector using a pre-trained sentence encoder, and each serverless function is represented by a precomputed embedding vector derived from its textual information (e.g., function name, docstring, and source code). Candidate functions are ranked by computing the cosine similarity between the query embedding and the function embeddings.


$\bullet$ \textit{Customized LLM-based variant method}: This baseline employs a prompt-based strategy in which LLMs directly generate intent summaries without extracting other information (e.g., platform, programming language, cloud service). The prompt consists of either the raw serverless function code or the task query, combined with an instruction to produce an intent summary. The resulting summaries are encoded into semantic vectors using a pre-trained sentence encoder, and recommendations are derived by computing cosine similarity. In contrast to \toolName, this method does not incorporate multi-level structural information extraction and candidate pruning.

\subsection{Experimental Settings}\label{sec:experimentsetting}

\noindent \textbf{Parameter Settings.} For \textbf{RQ1}, we compare \toolName with key-based method and embedding-based method. No parameter tuning is required. For \textbf{RQ2}, we compare \toolName with the customized LLM-based variant method, as both employ LLMs as a core component. We adopt ChatGPT-4o as the default LLM, owing to its wide adoption and demonstrated effectiveness in recent studies~\cite{yuan2024evaluating, lian2023configuration}. A critical hyperparameter in LLM-based generation is the \emph{temperature}, which controls the randomness of the output. To ensure reproducibility and deterministic behavior, we follow prior work~\cite{yin2024multitask, xu2024divlog, hadadi2024anomaly, chen2024automatic-paper7} and fix the temperature to 0 for all identical queries. In this RQ, we further evaluate \toolName against both key-based and embedding-based approaches, with a focus on recommendation latency.
For \textbf{RQ3}, we maintain the default temperature at 0 to analyze the non-determinism of LLMs under controlled conditions, and further vary it to 0.2 and 0.5 to examine the robustness of \toolName in other temperature settings. For \textbf{RQ4}, we investigate the generalization capability of \toolName across diverse leading LLMs beyond ChatGPT-4o. Specifically, we evaluate the top-ranked open-source model Llama 3.1 (405B) Instruct Turbo, the proprietary Gemini 2.0 Flash, and the widely adopted DeepSeek V3. 
Following the RQ2 setting, the temperature is fixed to 0 for all LLMs to ensure deterministic outputs.


\noindent \textbf{Evaluation Strategy and Metrics.} We evaluate \toolName against baselines by generating a ranked list of candidate functions for each task query. Due to the absence of publicly available query datasets for serverless function recommendations, we construct a synthetic query dataset to enable evaluation. 
This dataset requires that each task query be validated with the corresponding correct function, i.e., the \textit{ground-truth} function. To this end, we extract 150 task descriptions from the README files of the collected functions. When the README file of a function provides a specific and well-defined task description, we treat this description as a task query from the developer and designate the corresponding function as the \textit{ground truth}. The size of our dataset surpasses that of previous query datasets~\cite{husain2019codesearchnet, silavong2022senatus, chatterjee2009sniff, raghothaman2016swim, rahman2016rack, cambronero2019deep}. We then use the following evaluation metrics.

$\bullet$ \textit{Recall@$k$.} For each method, we compute Recall@$k$ by checking whether the \textit{ground-truth} function appears within the top-$k$ recommended results. The value of $k$ is adjusted to different settings, including 1, 5, 10, 15, and 20. This metric measures the coverage of recommendations for a given method, i.e., the proportion of task queries for which the \textit{ground-truth} function is recommended within the top-$k$ positions. Recall@$k$ ranges from 0 to 100\%, with higher values indicating stronger coverage and more effective function discovery.


$\bullet$ \textit{MRR@$k$.} Mean Reciprocal Rank (MRR) measures how early the \textit{ground-truth} function appears in the ranked list. For each task query, the reciprocal rank is defined as the inverse of the rank position of the \textit{ground-truth} function. If the \textit{ground-truth} function does not occur within the top-$k$ results, the reciprocal rank is set to zero. \textit{MRR@$k$} is then computed as the average reciprocal rank across all queries. \textit{MRR@$k$} = $\frac{1}{N} \sum_{i=1}^{N} \frac{1}{\text{rank}_i}$, where $N$ is the number of task queries and $\text{rank}_i$ denotes the position of  \textit{ground-truth} function for query $i$ in the top-$k$ list. The value of \textit{MRR@$k$} ranges from 0 to 1, with higher values indicating that correct functions tend to be ranked earlier.


$\bullet$ \textit{RecLatency.} We define the recommendation latency as the time elapsed from the submission of a task query to the completion of generating the recommendation list. \textit{RecLatency} is reported as the average latency across all task queries, reflecting the efficiency of function discovery.



\noindent \textbf{Experimental Repetitions.} To account for stochastic variability in experiments involving randomness, we follow established best practices~\cite{xu2024divlog, hadadi2024anomaly, wen2025scope} by repeating each experiment five times. Reported results correspond to the mean values of the evaluation metrics across these runs, thereby reducing the effect of random fluctuations and enhancing result reliability.


\noindent \textbf{Experimental Environment.} All experiments were conducted on an Ubuntu 18.04.4 LTS server equipped with an Intel Xeon (R) 4-core processor and 24 GiB of memory. 
Access to LLMs was obtained by their respective APIs.



\section{Evaluation Results}~\label{sec:results}


\subsection{Results of RQ1: Effectiveness of \toolName and Traditional Baselines}\label{sec:rq1result}

\begin{table*}[t]
\footnotesize
 \caption{(RQ1) \textit{Recall@$k$} and \textit{MRR@$k$} results for two traditional baseline methods and \toolName.}
 \label{tab:RQ1}
\begin{tabular}{l|l|c|c|c|c|c}
\hline
\textbf{Metrics} & \textbf{Methods} &  $k$ = 1 & $k$ = 5 & $k$ = 10 & $k$ = 15 & $k$ = 20\\
 \hline
    \multirow{6}{*}{\tabincell{l}{\textbf{Recall@$k$} \\(Improved \\percentage \\points)}} & Keyword-based method & 16.67\% & 32.00\% & 42.00\% & 48.67\% & 52.00\% \\ \cline{2-7}
    & Embedding-based method & 23.33\% & 50.67\% & 66.67\% & 74.00\% & 77.33\%  \\ \cline{2-7}
     &\tabincell{l}{\textbf{\toolName} \\ (vs. Keyword)} & \tabincell{c}{52.13\% \\ ($\uparrow$\textbf{35.47})} & \tabincell{c}{86.13\% \\ ($\uparrow$\textbf{54.13})} & \tabincell{c}{91.20\% \\ ($\uparrow$\textbf{49.20})} & \tabincell{c}{92.67\% \\ ($\uparrow$\textbf{44.00})}  & \tabincell{c}{92.93\% \\ ($\uparrow$\textbf{40.93})}  \\ \cline{2-7}
     & \tabincell{l}{(vs. Embedding)} & \tabincell{c}{($\uparrow$\textbf{28.80})} & \tabincell{c}{($\uparrow$\textbf{35.47})}  & \tabincell{c}{($\uparrow$\textbf{24.53})}  & \tabincell{c}{($\uparrow$\textbf{18.67})}  & \tabincell{c}{($\uparrow$\textbf{15.60})}  \\ \hline
     \multirow{6}{*}{\tabincell{l}{\textbf{MRR@$k$} \\(Improved\\ percentage)}} & Keyword-based method & 0.1667 & 0.2153 & 0.2291 & 0.2345 & 0.2363 \\ \cline{2-7}
     & Embedding-based method & 0.2333 & 0.3389 & 0.3593 & 0.3650 & 0.3669 \\ \cline{2-7}
     &\tabincell{l}{\textbf{\toolName} \\ (vs. Keyword)}& \tabincell{c}{0.5240 \\ ($\uparrow$\textbf{214.40\%})} & \tabincell{c}{0.6547 \\ ($\uparrow$\textbf{204.05\%})} & \tabincell{c}{0.6616 \\ ($\uparrow$\textbf{188.77\%})} & \tabincell{c}{0.6628 \\ ($\uparrow$\textbf{182.66\%})}  & \tabincell{c}{0.6629 \\ ($\uparrow$\textbf{180.50\%})}  \\ \cline{2-7}
     & \tabincell{l}{(vs. Embedding)} & \tabincell{c}{($\uparrow$\textbf{124.57\%})} & \tabincell{c}{($\uparrow$\textbf{93.19\%})}  & \tabincell{c}{($\uparrow$\textbf{84.15\%})}  & \tabincell{c}{($\uparrow$\textbf{81.58\%})}  & \tabincell{c}{($\uparrow$\textbf{80.70\%})}  \\ \hline
\end{tabular}    
\end{table*}


We explore the effectiveness of the semantic-enhanced representation introduced by \toolName against two traditional baselines. As shown in Table~\ref{tab:RQ1}, \toolName demonstrates significant advantages in \textit{Recall@k} and \textit{MRR@k}. We will introduce the results of each evaluation metric.
For \textit{Recall@k}, results show that \toolName consistently discovers the ground-truth function with higher coverage, highlighting its superior effectiveness. Specifically, \toolName achieves \textit{Recall@1} of 52.13\%, \textit{Recall@5} of 86.13\%, \textit{Recall@10} of 91.20\%, \textit{Recall@15} of 92.67\%, and \textit{Recall@20} of 92.93\%. In contrast, the keyword-based method yields lower values of 16.67\%, 32.00\%, 42.00\%, 48.67\%, and 52.00\% at the same cutoffs. The embedding-based method outperforms the keyword-based method but lags behind \toolName, with corresponding values of 23.33\%, 50.67\%, 66.67\%, 74.00\%, and 77.33\%. Relative to the keyword-based method, \toolName improves recall by 35.47, 54.13, 49.20, 44.00, and 40.93 percentage points across the five cutoffs. Compared to the embedding-based method, the improvements of \toolName are 28.80, 35.47, 24.53, 18.67, and 15.60 percentage points, respectively. 
In particular, \textit{Recall@1} of \toolName (52.13\%) not only far exceeds that of the keyword-based method at \textit{Recall@1}  (16.67\%), but also surpasses the result of the keyword-based method even at \textit{Recall@20} (52.00\%). This indicates that \toolName delivers higher discovery effectiveness.


For \textit{MRR@k}, \toolName demonstrates high ranking quality, as the ground-truth functions are more likely to appear earlier in the results. Specifically, \toolName achieves \textit{MRR@1} = 0.5240, \textit{MRR@5} = 0.6547, \textit{MRR@10} = 0.6616, \textit{MRR@15} = 0.6628, and \textit{MRR@20} = 0.6629. In contrast, the keyword-based method yields 0.1667, 0.2153, 0.2291, 0.2345, and 0.2363, while the embedding-based method attains 0.2333, 0.3389, 0.3593, 0.3650, and 0.3669, respectively. Relative to the keyword-based method, \toolName improves 214.40\%, 204.05\%, 188.77\%, 182.66\% and 180.50\%, and compared to the embedding-based method, by 124.57\%, 93.19\%, 84.15\%, 81.58\% and 80.70\%. All three approaches exhibit a noticeable increase from \textit{MRR@1} to \textit{MRR@5}, after which the values grow marginally and gradually converge. \toolName shows an improvement (0.5240 to 0.6547) compared with the keyword-based (0.1667 to 0.2153) and embedding-based (0.2333 to 0.3389) methods. Beyond \textit{MRR@5}, performance stabilizes, with the keyword-based method plateauing near 0.23, the embedding-based method near 0.36, and \toolName maintaining a substantially higher level around 0.66.

We further analyze the reasons behind the poor effectiveness of the two baselines. For the keyword-based method, two limitations are identified. First, the keyword match rate is inherently low. Among the 150 evaluated task queries, the average number of matched keywords for the ground-truth functions is only 7.92. This gap arises because queries generally employ everyday vocabulary and synonyms, whereas function implementations predominantly rely on technical jargon, library-specific naming conversions, and abbreviations. Such a lexical mismatch substantially reduces keyword coverage, even after applying standard preprocessing techniques such as stop-word removal and stemming. Second, this method fails to capture the discriminative features that characterize functional intent. We analyze keyword frequency and reveal that high-frequency keywords such as ``exampl'' (106 occurrences), ``use'' (88 occurrences), and ``demonstr'' (65 occurrences) dominate the matching process. These keywords are generic and semantically uninformative, contributing little to distinguishing the task goal from the code functionality. As a result, the keyword-based method struggles to identify the ground truth functions, leading to low recall and MRR scores. Overall, the keyword-based method lacks the capacity to model semantic relevance at a higher level, making it relatively ineffective.

For the embedding-based method, several limitations are observed. 
First, embeddings learned by this method primarily rely on surface vocabularies. Due to the inherent inconsistency between task queries and function implementations, it introduces semantic bias and results in weak similarity and suboptimal recommendations. Second, embeddings generated from general-purpose language models are not optimized for code-specific semantics. They fail to capture functional characteristics of program logic, such as serverless platform usage patterns, languages, or task goals. As a result, the semantic gap between how functionality is expressed in natural language task queries and how it is encoded in function implementations remains unbridged. Finally, embeddings tend to dilute task-specific signals by overemphasizing frequent but non-discriminative terms, leading to insufficient differentiation between functions with similar lexical content but distinct functional purposes. 
Overall, while the embedding-based method can partially capture task query intent, it struggles to represent code functionality intent due to the abstract and heterogeneous nature of code expression. Thus, this baseline cannot establish strong alignment with task queries.

By contrast, \toolName addresses these limitations by introducing a semantic-enhanced representation that models richer relationships between task queries and function implementations at a deeper semantic level. It explicitly mines the underlying functionality intent from code structures.



\finding{
For recommendation quality, \toolName achieves \textit{Recall@1} of 52.13\%, \textit{Recall@5} of 86.13\%, \textit{Recall@10} of 91.20\%, \textit{Recall@15} of 92.67\%, and \textit{Recall@20} of 92.93\%, surpassing the state-of-the-art embedding-based baseline by 28.80, 35.47, 24.53, 18.67, and 15.60 percentage points, respectively. For ranking quality, \toolName attains \textit{MRR@1} of 0.5240, \textit{MRR@5} of 0.6547, \textit{MRR@10} of 0.6616, \textit{MRR@15} of 0.6628, and \textit{MRR@20} of 0.6629, corresponding to relative improvements of 124.57\%, 93.19\%, 84.15\%, 81.58\%, and 80.70\% over the same baseline.
These results suggest the effectiveness of the core semantic-enhanced representation design of \toolName.

}

\subsection{Results of RQ2: Advantage of Multi-Level Structural Pruning}



\begin{figure}[htbp]
    \centering
    \begin{minipage}[t]{0.49\textwidth}
        \centering
        \includegraphics[width=\textwidth]{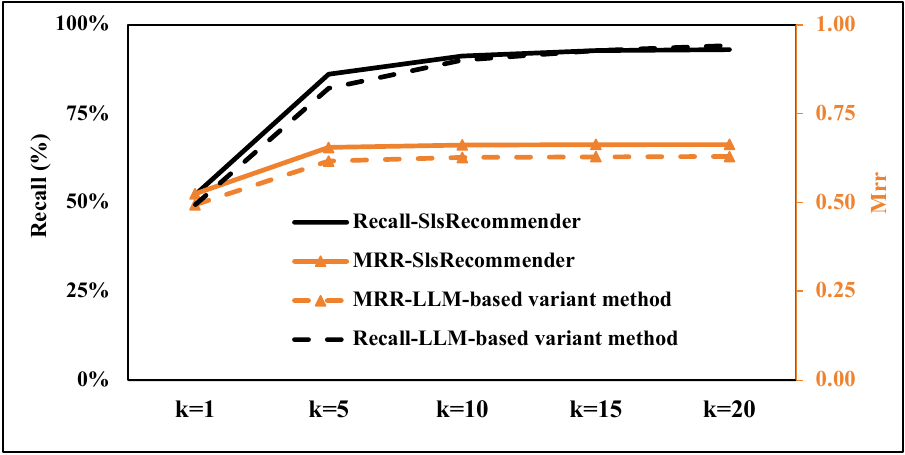}
        \caption{(RQ2) \textit{Recall@k} and \textit{MRR@k} results of \toolName and the customized LLM-based variant.}
        \label{fig:Baseline3VSOur}
    \end{minipage}
    \hfill
    \begin{minipage}[t]{0.49\textwidth}
        \centering
        \includegraphics[width=\textwidth]{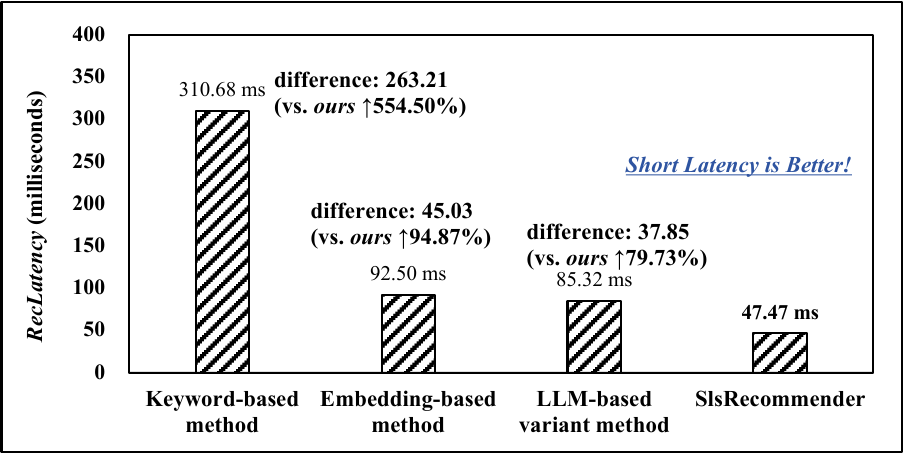}
        \caption{(RQ2) Recommendation latency comparisons between \toolName and three baselines.}
        \label{fig:RecLatencyData}
    \end{minipage}
\end{figure}

To assess the additional benefits of \toolName, we compare it with a customized LLM-based variant that omits the multi-level structural pruning step. As shown in Fig.~\ref{fig:Baseline3VSOur}, \toolName achieves recall and MRR scores comparable to those of the variant across different $k$ values. This indicates that both methods are similarly effective in retrieving relevant functions. The performance of the customized variant is expected, as it still adheres to the same principle of aligning task queries with function-level intent, thereby enabling the identification of semantically related candidates. The comparable results of \toolName confirm that multi-level pruning retains the vast majority of relevant candidates, validating its effectiveness without compromising accuracy.


We further evaluate the recommendation latency of \toolName against three baselines. As shown in Fig.~\ref{fig:RecLatencyData}, \toolName achieves an average latency of 47.47 ms, significantly faster than the customized LLM-based variant (85.32 ms), the keyword-based method (310.68 ms), and the embedding-based method (92.50 ms). In absolute terms, the customized LLM-based variant is slower by 37.85 ms (79.73\% longer than \toolName), the keyword-based method by 263.21 ms (554.50\% longer), and the embedding-based method by 45.03 ms (94.87\% longer). From the perspective of \toolName, it reduces the recommendation latency by 44.36\%, 84.72\%, and 48.68\% compared with the three baselines, demonstrating clear efficiency advantages. The major bottleneck for baselines is that they compute similarity scores exhaustively against all candidate functions, incurring considerable overhead. By contrast, \toolName leverages a multi-level pruning strategy grounded in structural information (e.g., serverless platforms, programming languages, and cloud services), which narrows the search space. We further analyze the candidate set size after applying our multi-level pruning for each task query. On average, fewer than 40\% of functions remain for subsequent similarity computation, thereby substantially reducing the search space and improving recommendation efficiency.
These results show that \toolName not only achieves effective recommendation results but also delivers significantly lower recommendation latency, underscoring its practical applicability.

\finding{
\toolName preserves most relevant candidates even with multi-level pruning, demonstrating the effectiveness of its pruning strategy. Moreover, it achieves an average recommendation latency of 47.47 ms, representing reductions of 44.36\%, 84.72\%, and 48.68\% compared with the LLM-based variant, the keyword-based method, and the embedding-based method, respectively. These results highlight the recommendation efficiency of \toolName.

}


\subsection{Results of RQ3: Impact of Non-determinism on \toolName}

\begin{table*}[t]
\footnotesize
 \caption{(RQ3) \textit{Recall@k} and \textit{MRR@k} results of \toolName across five repetitions.}
    \label{tab:RQ3}
    \begin{tabular}{c|l|c|c|c|c|c|c}
    \hline
     $k$ values  & Temperature &  Repetition 1 & Repetition 2 & Repetition 3 & Repetition 4 &  Repetition 5 & \textbf{Mean} \\
    \hline
    \multirow{3}{*} & temperature=0 &\tabincell{c} {49.33\% /\\0.4933} & \tabincell{c} {52.67\% /\\0.5400} & \tabincell{c}{56.67\% /\\0.5667} & \tabincell{c}{50.67\% /\\0.5067} & \tabincell{c}{51.33\% /\\0.5133} & \tabincell{c}{52.13\% /\\0.5240} \\ \cline{2-8}
    \tabincell{c}{$k$ = 1} & temperature=0.2 & \tabincell{c}{54.00\% /\\0.5400} & \tabincell{c}{51.33\% /\\0.5133} & \tabincell{c}{50.00\% /\\0.5000} & \tabincell{c}{49.33\% /\\0.4933} & \tabincell{c}{54.67\% /\\0.5467} & \tabincell{c}{51.87\% /\\0.5187} \\ \cline{2-8}
    \tabincell{c}{(recall / \\ MRR)} & temperature=0.5  & \tabincell{c}{48.00\% /\\0.4800} & \tabincell{c}{44.67\% /\\0.4467} & \tabincell{c}{49.33\% /\\0.4933} & \tabincell{c}{47.33\% /\\0.4733} & \tabincell{c}{46.67\% /\\0.4667} & \tabincell{c}{47.20\% /\\0.4720} \\ \hline

    \multirow{3}{*} & temperature=0   & \tabincell{c}{83.33\% /\\0.6219} & \tabincell{c}{87.33\% /\\0.6633} & \tabincell{c}{88.00\% /\\0.6893} & \tabincell{c}{86.67\% /\\0.6496} & \tabincell{c}{85.33\% /\\0.6494} & \tabincell{c}{86.13\% /\\0.6547} \\ \cline{2-8}
    $k$ = 5 & temperature=0.2   & \tabincell{c}{84.00\% /\\0.6471} & \tabincell{c}{82.00\% /\\0.6287} & \tabincell{c}{83.33\% /\\0.6257} & \tabincell{c}{85.33\% /\\0.6297} & \tabincell{c}{84.67\% /\\0.6583} & \tabincell{c}{83.87\% /\\0.6379} \\ \cline{2-8}
    \tabincell{c}{(recall / \\ MRR)} & temperature=0.5 & \tabincell{c}{79.33\% /\\0.5958} & \tabincell{c}{82.00\% /\\0.5926} & \tabincell{c}{81.33\% /\\0.6210} & \tabincell{c}{80.00\% /\\0.5953} & \tabincell{c}{86.67\% /\\0.6064} & \tabincell{c}{81.87\% /\\0.6022} \\ \hline

    \multirow{3}{*} & temperature=0  & \tabincell{c}{86.67\% /\\0.6273} & \tabincell{c}{90.67\% /\\0.6676} & \tabincell{c}{92.67\% /\\0.6958} & \tabincell{c}{93.33\% /\\0.6581} & \tabincell{c}{92.67\% /\\0.6593} & \tabincell{c}{91.20\% /\\0.6616} \\ \cline{2-8}
    $k$ = 10 & temperature=0.2 & \tabincell{c}{92.00\% /\\0.6581} & \tabincell{c}{90.67\% /\\0.6407} & \tabincell{c}{92.00\% /\\0.6373} & \tabincell{c}{90.67\% /\\0.6375} & \tabincell{c}{92.00\% /\\0.6687} & \tabincell{c}{91.47\% /\\0.6485} \\ \cline{2-8}
    \tabincell{c}{(recall / \\ MRR)} & temperature=0.5 & \tabincell{c}{88.67\% /\\0.6079} & \tabincell{c}{90.00\% /\\0.6034} & \tabincell{c}{88.67\% /\\0.6316} & \tabincell{c}{86.00\% /\\0.6042} & \tabincell{c}{90.67\% /\\0.6124} & \tabincell{c}{88.80\% /\\0.6119} \\\hline

     \multirow{3}{*} & temperature=0 & \tabincell{c}{89.33\% /\\0.6293} & \tabincell{c}{92.67\% /\\0.6689} & \tabincell{c}{94.00\% /\\0.6969} & \tabincell{c}{94.00\% /\\0.6587} & \tabincell{c}{93.33\% /\\0.6598} & \tabincell{c}{92.67\% /\\0.6628} \\ \cline{2-8}
    $k$ = 15 & temperature=0.2 & \tabincell{c}{94.00\% /\\0.6598} & \tabincell{c}{91.33\% /\\0.6412} & \tabincell{c}{93.33\% /\\0.6384} & \tabincell{c}{92.67\% /\\0.6389} & \tabincell{c}{92.67\% /\\0.6691} & \tabincell{c}{92.80\% /\\0.6495} \\ \cline{2-8}
    \tabincell{c}{(recall / \\ MRR)} & temperature=0.5 & \tabincell{c}{90.67\% /\\0.6096} & \tabincell{c}{90.67\% /\\0.6040} & \tabincell{c}{90.67\% /\\0.6333} & \tabincell{c}{88.00\% /\\0.6060} & \tabincell{c}{91.33\% /\\0.6129} & \tabincell{c}{90.27\% /\\0.6132} \\ \hline

     \multirow{5}{*} & temperature=0 & \tabincell{c}{89.33\% /\\0.6293} & \tabincell{c}{92.67\% /\\0.6690} & \tabincell{c}{94.67\% /\\0.6972} & \tabincell{c}{94.00\% /\\0.6587} & \tabincell{c}{94.00\% /\\0.6602} & \tabincell{c}{92.93\% /\\0.6629} \\ \cline{2-8}
   $k$ = 20 & temperature=0.2 & \tabincell{c}{94.00\% /\\0.6598} & \tabincell{c}{91.33\% /\\0.6412} & \tabincell{c}{94.00\% /\\0.6388} & \tabincell{c}{93.33\% /\\0.6394} & \tabincell{c}{95.33\% /\\0.6705} & \tabincell{c}{93.60\% /\\0.6499} \\ \cline{2-8}
   \tabincell{c}{(recall / \\ MRR)} & temperature=0.5 & \tabincell{c}{91.33\% /\\0.6100} & \tabincell{c}{90.67\% /\\0.6040} & \tabincell{c}{91.33\% /\\0.6337} & \tabincell{c}{88.67\% /\\0.6063} & \tabincell{c}{93.33\% /\\0.6140} & \tabincell{c}{91.07\% /\\0.6136} \\ \hline

\end{tabular}
\end{table*}

We investigate the impact of LLM non-determinism on evaluation results under a default temperature setting of 0. As described in Section~\ref{sec:experimentsetting}, each experiment is repeated five times, and we report \textit{Recall@k} and \textit{MRR@k} at different cutoff values (Table~\ref{tab:RQ3}). Results show that while LLM non-determinism introduces fluctuations, its effect is relatively limited, with recall varying within approximately 5 percentage points and \textit{MRR} differing by no more than 0.06 across trials in different $k$. Under temperature 0, \toolName consistently shows strong effectiveness. Specifically, \textit{Recall@1} ranges from 49.33\% to 56.67\%, \textit{Recall@5} from 83.33\% to 88.00\%, \textit{Recall@10} from 86.67\% to 93.33\%, \textit{Recall@15} from 89.33\% to 94.00\%, and \textit{Recall@20} from 89.33\% to 94.67\%. Particularly, even the lowest values at each cutoff (e.g., 83.33\% at \textit{Recall@5} and 86.67\% at \textit{Recall@10}) still surpass both keyword-based and embedding-based baselines. A similar trend is observed for \textit{MRR@k}, further confirming that the observed superiority of \toolName is robust to the non-determinism of LLMs.


To further assess the robustness of \toolName, we analyze the experimental results at temperatures 0.2 and 0.5. As shown in Table~\ref{tab:RQ3}, there is a relatively minor impact. When the temperature is 0.2, the recall variance is about 3\% and the MRR variance is about 0.03 across all repetitions in all $k$ values. When the temperature is 0.5, the recall variance is about 5\% and the MRR variance is about 0.06 across all repetitions. From the results, the temperature of 0.5 has more randomness than that of 0.2. It is reasonable that the higher the temperature setting, the greater the randomness. 
Specifically, when the temperature is set to 0.2, \textit{Recall@1} ranges from 49.33\% to 54.67\%, \textit{Recall@5} from 82.00\% to 85.33\%, \textit{Recall@10} from 90.67\% to 92.00\%, \textit{Recall@15} from 91.33\% to 94.00\%, and \textit{Recall@20} from 91.33\% to 95.33\%, with respective differences of 5.33, 3.33, 1.33, 2.67, and 4.00 percentage points. The corresponding MRR values are 0.4933–0.5467 for \textit{MRR@1}, 0.6257–0.6583 for \textit{MRR@5}, 0.6373–0.6686 for \textit{MRR@10}, 0.6384–0.6691 for \textit{MRR@15}, and 0.6388–0.6705 for \textit{MRR@20}, with variations of 0.0533, 0.0327, 0.0313, 0.0307, and 0.0317, respectively. At temperature = 0.5, the fluctuations are slightly larger. \textit{Recall@1} ranges from 44.67\% to 49.33\%, \textit{Recall@5} from 79.33\% to 86.67\%, \textit{Recall@10} from 86.00\% to 90.67\%, \textit{Recall@15} from 88.00\% to 91.33\%, and \textit{Recall@20} from 88.67\% to 93.33\%, with respective differences of 4.67, 7.33, 4.67, 3.33, and 4.67 percentage points. For MRR, \textit{MRR@1} ranges from 0.4467 to 0.4933, \textit{MRR@5} from 0.5926 to 0.6210, \textit{MRR@10} from 0.6034 to 0.6316, \textit{MRR@15} from 0.6040 to 0.6333, and \textit{MRR@20} from 0.6040 to 0.6337, with respective differences of 0.0733, 0.0674, 0.0685, 0.0676, and 0.0680. 
Despite these variations, the effectiveness of \toolName remains consistently high across all evaluation metrics and is comparable to the default setting (temperature = 0). 
Particularly, even the lowest observed values across all metrics still outperform the keyword-based and embedding-based baselines. These experimental results confirm the strong robustness of \toolName.

\finding{The non-determinism of LLMs does not affect the validity of our conclusions.}



\subsection{Results of RQ4: Generalization Capability of \toolName}

\begin{table*}[t]
\footnotesize
\caption{(RQ4) \textit{Recall@k} and \textit{MRR@k} results about \toolName using various LLMs.}
    \label{tab:RQ4}
    \begin{tabular}{l|c|c|c|c|c}
    \hline
      \textbf{LLMs} & k = 1 & k = 5 &  k = 10 & k = 15 & k = 20 \\
    \hline
    ChatGPT-4o & 52.13\% / 0.5240 & 86.13\% / 0.6547 & 91.20\% / 0.6616 & 92.67\% / 0.6628 & 92.93\% / 0.6629 \\ \hline
     \tabincell{l}{Llama 3.1 (405B) \\Instruct Turbo}     & 48.67\% / 0.4867 & 81.87\% / 0.6109 & 86.40\% / 0.6171 & 87.87\% / 0.6182 & 88.40\% / 0.6185  \\ \hline
     Gemini 2.0 Flash    & 42.40\% / 0.4240 & 76.80\% / 0.5525 & 81.73\% / 0.5594 & 82.13\% / 0.5597 & 83.60\% / 0.5605 \\ \hline
     DeepSeek V3  & 44.27\% / 0.4427 & 74.67\% / 0.5631 & 81.20\% / 0.5717 & 83.47\% / 0.5735 & 84.40\% / 0.5740   \\ \hline
\end{tabular}
\end{table*}

To explore the generalization capability of \toolName, we further experiment with three additional LLMs: Llama 3.1 (405B) Instruct Turbo, Gemini 2.0 Flash, and DeepSeek V3. As shown in Fig.~\ref{tab:RQ4}, \toolName consistently delivers high effectiveness across all models, achieving \textit{Recall@20} above 80\% and $MRR@20$ above 0.55. Specifically, with the Llama 3.1 (405B) Instruct Turbo, \toolName achieves \textit{Recall@1} = 48.67\%, \textit{Recall@5} = 81.87\%, \textit{Recall@10} = 86.40\%, \textit{Recall@15} = 87.87\%, and \textit{Recall@20} = 88.40\%, with corresponding \textit{MRR@1} = 0.4867, \textit{MRR@5} = 0.6109, \textit{MRR@10} = 0.6171, \textit{MRR@15} = 0.6182, and \textit{MRR@20} = 0.6185. For the Gemini 2.0 Flash, the results are \textit{Recall@1} = 42.40\%, \textit{Recall@5} = 76.80\%, \textit{Recall@10} = 81.73\%, \textit{Recall@15} = 82.13\%, and \textit{Recall@20} = 83.60\%, with \textit{MRR@1} = 0.4240, \textit{MRR@5} = 0.5525, \textit{MRR@10} = 0.5594, \textit{MRR@15} = 0.5597, and \textit{MRR@20} = 0.5605. With the DeepSeek V3, \toolName achieves \textit{Recall@1} = 44.27\%, \textit{Recall@5} = 74.67\%, \textit{Recall@10} = 81.20\%, \textit{Recall@15} = 83.47\%, and \textit{Recall@20} = 84.40\%, with \textit{MRR@1} = 0.4427, \textit{MRR@5} = 0.5631, \textit{MRR@10} = 0.5717, \textit{MRR@15} = 0.5735, and \textit{MRR@20} = 0.5740. Overall, the better effectiveness is observed with ChatGPT-4o and Llama 3.1 (405B) Instruct Turbo, while Gemini 2.0 flash model and DeepSeek V3 achieve slightly lower metrics but still maintain high \textit{Recall} values (83.60\% and 84.40\% at $k$ = 20).


We further investigate the reasons why Gemini 2.0 Flash and DeepSeek V3 exhibit comparatively weaker performance. Our analysis indicates that both models exhibit weaknesses in structural knowledge extraction, particularly in recognizing serverless platforms. On average, 7–9 task queries per evaluation round yield incorrect results due to platform misclassification, which directly lowers recall and MRR. This suggests that the inferior performance of Gemini 2.0 Flash and DeepSeek V3 arises from inherent limitations in structured metadata inference.


\finding{
\toolName exhibits generalization capability, consistently yielding effective results across diverse LLMs. \toolName achieves better effectiveness when combined with ChatGPT-4o and Llama 3.1 (405B) Instruct Turbo.

}


\section{Threats to Validity}\label{sec:discussion}

\textbf{Evaluation Query Data.} Since no publicly available query datasets exist for serverless function recommendation, we construct a dedicated dataset for evaluation. A potential threat to validity arises from the absence of real developer-requirement queries. To mitigate this issue, we systematically analyze a curated repository of reusable serverless functions and identify cases where original function descriptions can be reasonably reinterpreted as developer queries. In these cases, the corresponding functions are treated as ground-truth recommendations. This strategy ensures that the constructed dataset remains both verifiable and representative of practical usage scenarios.


\noindent \textbf{Data Leakage Concerns.} A potential threat is data leakage, since LLMs are pretrained on large-scale corpora that may include open-source code and serverless platform examples. As a result, models may have been exposed to content resembling the functions or queries used in our evaluation. We regard the impact on our results as negligible for two reasons. First, our function repository and evaluation queries are manually curated and refined into task-oriented requirements, reducing the likelihood of direct overlap with pretraining data. Second, our study focuses on structured information extraction, which requires contextual reasoning rather than rote memorization. Thus, even if partial overlap exists, it is unlikely to compromise the validity of our conclusions.

\section{Related Work}\label{sec:relatedwork}

\noindent \textbf{Serverless Computing.} The rapid adoption of serverless computing has spurred substantial research interest, leading to broad studies, such as performance characterization and testing~\cite{kim2019functionbench, yu2020characterizing, maissen2020faasdom, wen2025scope}, performance optimization~\cite{eismann2021sizeless, wen2023FaaSlight}, usage characterization~\cite{eismann2021state, jangda2019formal, shahrad2020serverless}, and development challenges~\cite{Wen21challenges}. For example, FunctionBench~\cite{kim2019functionbench} provided a benchmark suite for evaluating serverless function performance across multiple platforms. SCOPE~\cite{wen2025scope} introduced a performance testing method to address performance variance problems in function executions. To improve resource provisioning, Sizeless~\cite{eismann2021sizeless} was presented as a predictive approach that selects optimal resource configurations. From a usage perspective, Eismann \textit{et al.}\cite{eismann2021state} conducted an empirical analysis of serverless functions, providing insights into their structural and operational properties. Wen \textit{et al.}\cite{Wen21challenges} identified 36 challenges in function development, underscoring key obstacles to developer productivity. Despite these advances, the challenge of enabling effective serverless function reuse remains unresolved. This paper fills this gap by proposing \toolName.

\noindent \textbf{Code Recommendation Studies.} This paper centers on function code recommendations. Prior research closely related to our study can be grouped into two directions: natural language–to–code recommendation and code-to-code recommendation. First, natural language–to–code recommendation has investigated methods for discovering or generating code snippets from natural language queries~\cite{chatterjee2009sniff, cambronero2019deep, gu2018deep, sachdev2018retrieval, gao2023know}. For instance, Que2Code~\cite{gao2023know} identified relevant code snippets for queries by leveraging Stack Overflow posts. However, it is limited to fragmentary code solutions and cannot deliver complete function implementations. SNIFF~\cite{chatterjee2009sniff} supported keyword-based matching to enhance code discovery from queries. We compare this method with \toolName in Section~\ref{sec:rq1result}. \toolName achieves substantial recall improvements of 35.47–54.13 percentage points. Recent approaches have investigated mapping between queries and source code~\cite{cambronero2019deep, gu2018deep, sachdev2018retrieval}. For example, NCS~\cite{gu2018deep} and UNIF~\cite{cambronero2019deep} leveraged neural networks to jointly embed source code and natural language queries into a shared space. We also compare \toolName with UNIF, one of the most effective text-to-code recommendation methods~\cite{cambronero2019deep}. \toolName improves recall by 15.60–35.47 percentage points over UNIF. The reasons for the limited effectiveness of these methods are discussed in Section~\ref{sec:rq1result}.

A separate line of research has explored code recommendation from code queries~\cite{kim2018facoy, luan2019aroma, silavong2022senatus, nguyen2020code}. 
AROMA~\cite{luan2019aroma} indexed Java methods, clustered candidates, and intersected retrieval results to produce concise code snippets that captured usage patterns. GROUM~\cite{nguyen2020code} modeled code as graphs to represent method calls and control-flow dependencies, enabling fine-grained pattern matching for Android applications. While effective, these approaches target code fragment similarity, whereas our work addresses recommending reusable serverless functions from natural language requirements.

\section{Conclusion}\label{sec:conclusion}


In this paper, we presented \toolName, the first LLM-powered framework for serverless function reuse. By constructing a reusable function repository and employing prompt-engineered semantic enhancement, \toolName bridged the gap between natural language queries and heterogeneous function implementations. Its intent-aware function discovery, enhanced by multi-level pruning and similarity ranking, enabled matching of task queries to reusable functions. Evaluation on 110 task queries showed that \toolName significantly outperformed the state-of-the-art method, achieving \textit{Recall@1} and \textit{Recall@10} gains of 28.80 and 24.53 percentage points, respectively. Additional experiments with ChatGPT-4o, Llama 3.1 (405B) Instruct Turbo, Gemini 2.0 Flash, and DeepSeek V3 demonstrate the generalization of \toolName, confirming its consistent effectiveness across models.


\bibliographystyle{ACM-Reference-Format}
\bibliography{main}

\end{document}